\documentclass[journal]{IEEEtran}
\ifCLASSINFOpdf
\else
\fi
\usepackage{graphicx}
\usepackage{cite}
\usepackage{color}
\usepackage{indentfirst}
\usepackage{booktabs}
\usepackage{tabularx}  
\usepackage{tabu}  
\usepackage{makecell}
\usepackage{multirow} 
\usepackage{upgreek}
\usepackage{siunitx}
\usepackage{amssymb}
\usepackage{amsmath}
\begin{document}
%
\title{Interference Factors and Compensation Methods when Using Infrared Thermography for Temperature Measurement: A Review}

\author{Dong Pan, Tan Mo, Zhaohui Jiang, \IEEEmembership{Member, IEEE}, Yuxia Duan, Xavier Maldague, and Weihua Gui
\thanks{This work was supported by the Young Scientists Fund of the National Natural Science Foundation of China under Grant 62303491; in part by the Project of State Key Laboratory of Precision Manufacturing for Extreme Service Performance (Grant No. ZZYJKT2023-14); in part by the Science and Technology Innovation Program of Hunan Province(Grant No.2024RC1007) and in part by the Open Research Project of the State Key Laboratory of Industrial Control Technology (Grant No. ICT2024B05).}
\thanks{D. Pan is with the State Key Laboratory of Precision Manufacturing for Extreme Service Performance and the School of Automation, Central South University, Changsha 410083, China (e-mail: pandong@csu.edu.cn).}
\thanks{T. Mo, Z. H. Jiang and W. H. Gui are with the School of Automation, Central South University, Changsha 410083, China (e-mail: 244611093@csu.edu.cn; jzh0903@csu.edu.cn; gwh@csu.edu.cn).}
\thanks{Y. X. Duan is with the School of Physics, Central South University, Changsha 410083, China (yuxia.duan@csu.edu.cn).}
\thanks{Xavier Maldague is with the Department of Electrical and Computing
Engineering, Université Laval, Québec, QC G1V 0B1, Canada (e-mail:
xavier.maldague@gel.ulaval.ca).}
}

\maketitle

\begin{abstract}
Infrared thermography (IRT) is a widely used temperature measurement technology, but it faces the problem of measurement errors under interference factors. This paper attempts to summarize the common interference factors and temperature compensation methods when applying IRT. According to the source of factors affecting the infrared temperature measurement accuracy, the interference factors are divided into three categories: factors from the external environment, factors from the measured object, and factors from the infrared thermal imager itself. At the same time, the existing compensation methods are classified into three categories: Mechanism Modeling-based Compensation method (MMC), Data-Driven Compensation method (DDC), and Mechanism and Data jointly driven Compensation method (MDC). Furthermore, we discuss the problems existing in the temperature compensation methods and future research directions, aiming to provide some references for researchers in academia and industry when using IRT technology for temperature measurement.
\end{abstract}


%
\IEEEpeerreviewmaketitle

\section{Introduction}
%
%
%
%
\IEEEPARstart{I}{nfrared} thermography (IRT) is a powerful measurement technology with the advantages of non-contact, in-situ detection, no harmful radiation, and providing two-dimensional temperature field \cite{ref1}. IRT applications can be seen in many industries such as power electronics, aerospace, machinery, metallurgy, medical, construction, agriculture, archaeology, nuclear energy, and military \cite{ref2, ref3, ref4, ref5, ref6, ref7, ref8, ref9, ref10, ref11}. Temperature measurement and non-destructive testing are two common applications of IRT technology \cite{ref1}. Infrared temperature measurement (ITM) focuses more on the measurement accuracy of IRT, while non-destructive testing (NDT) focuses more on the temperature field distribution and does not pay much attention to the accuracy of the temperature measurement results \cite{ref12, ref13, ref14}. This paper primarily concentrates on ITM, investigating the factors that interfere with its accuracy and exploring corresponding compensation methods, while excluding any investigation into NDT.

Temperature is a parameter that must be measured in many industrial processes or scientific experiments for the purpose of temperature control, quality assessment, and safe production \cite{ref15, ref16, ref17}. Thanks to the advantages of IRT technology, it has been applied in many industrial processes, providing important temperature information for temperature control, energy consumption analysis, safety monitoring, etc. While the advantages of IRT contribute to its widespread application, this does not imply that ITM is without flaws. Given that IRT operates as a non-contact measurement technique, both the characteristics of the object being measured and the environmental conditions along the optical path between the measured object and the infrared thermal imager can significantly influence ITM accuracy when employing IRT \cite{ref18, ref19, ref20}. If temperature measurement errors are excessively large, they may severely compromise subsequent tasks related to monitoring, control, optimization, and other temperature-dependent operations, potentially leading to significant consequences. To ensure both accuracy and reliability in ITM results, it is imperative to implement compensatory measures for affected temperature measurements.

In fact, when using IRT technology for temperature measurement, various interference factors often arise in the detection environment due to the characteristics of the measured object, the production process, or limitations inherent to the installation environment. These factors include dust, water vapor, ambient temperature, humidity levels, and measurement distance, etc. Each of these factors can introduce errors and present significant challenges to ensuring ITM accuracy \cite{ref21, ref22, ref23, ref24, ref25, ref26}. For example, during the blast furnace ironmaking process, there is randomly distributed dust near the taphole, which bring a great challenge to the measurement of the molten iron temperature at the taphole based on infrared vision \cite{ref27}. The same dust influence can be found in the sintering process. The dust raised by the falling sintered ore greatly affects the detection of the cross-section temperature of the sintered ore \cite{ref28}. The aforementioned examples primarily highlight dust as an example to introduce the influence of interference factors on the accuracy of ITM, not to mention the influence of other interference factors on ITM.

How to ensure the temperature measurement accuracy is a huge challenge in the application of IRT technology. Over the past few decades, many scholars have carried out a lot of research on overcoming the influence of interference factors and ensuring the accuracy of ITM, and have achieved encouraging results in the laboratory and some industrial scenes \cite{ref29, ref30, ref31, ref32}. This paper focuses on the temperature measurement errors and compensation methods faced in the application of IRT, and it is assumed that commercial infrared temperature measurement instruments have completed calibration and correction before leaving the factory. The term ‘compensation’ is employed herein to indicate that when using infrared temperature measurement instruments, various interference factors will make it difficult for the accuracy of instruments to reach the measurement accuracy before leaving the factory, and the original temperature needs to be compensated to reduce the error. For this reason, compensation methods for different interference factors have developed to minimize errors induced by interference factors on ITM results, thereby striving to maintain the accuracy of ITM instruments as much as possible.

At present, some scholars have reviewed and analyzed IRT technology from the perspectives of electrical, mechanical, medical, and other applications. Rubén \emph{et al.} reviewed the research on IRT in temperature measurement and non-destructive testing, and introduced the advantages and principles of infrared thermal imaging as well as commonly used data processing techniques \cite{ref1}. Roque \emph{et al.} introduced the application of IRT in electrical, mechanical, and other applications, and focused on the application of IRT in induction motors \cite{ref33}. Bagavathiappan \emph{et al.} discussed in detail the research progress of IRT as a non-contact, non-invasive condition monitoring tool in the fields of machinery, equipment, and aerospace, and reviewed the data analysis technology related to IRT \cite{ref34}. Emilios \emph{et al.} reviewed the application of infrared radiation temperature measurement methods in temperature monitoring during metal turning, drilling, and milling, and discussed and evaluated the main advantages and limitations of thermocouples and infrared temperature measurement methods \cite{ref35}. Ayca \emph{et al.} reviewed in detail the research on using IRT to investigate abnormalities in building envelope structures, and classified the measurement methods, analysis schemes, etc. to highlight the potential of IRT in providing energy-saving solutions in building envelope detection \cite{ref36}. References \cite{ref37, ref38} summarized the research progress of IRT technology in animal surface temperature measurement and its applications in early diagnosis of diseases, monitoring of animal stress levels, estrus and ovulation identification, and discussed future research directions. Angeliki \emph{et al.} reviewed the research progress of passive infrared thermal imagers and active infrared thermal imagers in building diagnosis, and discussed the future development trend of IRT in building diagnosis \cite{ref39}. Lahiri \emph{et al.} focused on the research progress of IRT in medical, and introduced the basic principles of IRT and its application in various medical fields such as breast cancer, diabetes, and skin diseases \cite{ref40}. Reference \cite{ref41} focused on the image processing algorithms of infrared thermal imagers such as image non-uniformity correction, denoising and image pseudo-color enhancement, and analyzed the advantages and disadvantages of blind pixel detection and compensation, temperature measurement, target detection and tracking technologies.

The aforementioned reviews play an important role in understanding the principles of infrared temperature measurement and its application in different fields. However, there is still a lack of comprehensive summary and analysis of the interference factors and compensation methods for infrared temperature measurement according to Google Scholar and other available databases. The academic and industrial communities have conducted many theoretical research and practical exploration on the interference factors and corresponding temperature compensation methods in ITM, including mechanism modeling-based compensation method \cite{ref42}, data-driven compensation method \cite{ref43}, and mechanism and data jointly driven compensation method\cite{ref44}, etc. Although these studies have yielded some satisfactory results, using IRT to accurately obtain temperature information in complex scenes is not a simple task due to the complexity of the detection environment in which the infrared instrument is located, the diversity of the objects being measured, and the performance limitation of the infrared detector itself. Up to now, there is no review of interference factors and infrared temperature compensation methods.

To advance the development of IRT for temperature measurement, broaden the application scenarios of IRT, ensure the ITM accuracy in different scenarios, it is essential to provide scholars and engineers interested in IRT application in complex scenarios with a systematic and comprehensive understanding of interference factors and compensation methods for infrared temperature measurement. To this end, this article briefly reviews the principles of ITM, focuses on analyzing the interference factors that affect the accuracy of ITM, summarizes the existing ITM compensation methods, discusses the problems existing in ITM compensation methods, and offers some insights into the potential future direction for ITM compensation methods.

The rest of the paper is organized as follows. Section II briefly reviews the principle of IRT. Section III analyzes the interference factors that affect ITM. Section IV classifies the existing ITM compensation methods. Section V analyzes the main issues of ITM compensation methods and discusses some future research directions. Finally, Section VI concludes the work.

\section{Temperature Measurement Principle of IRT}
The basic principle of infrared temperature measurement is based on radiation thermometry theory, that is, any object with a surface temperature higher than absolute zero (-273.15℃) will emit infrared radiation, and there is a functional relationship between the infrared radiation and the surface temperature of the object. The higher the surface temperature of the object, the greater the intensity of the infrared radiation emitted \cite{ref45, ref46}.
\subsection{Basic Laws of Infrared Radiation}
Kirchhoff’s law \cite{ref47} states that the radiant power emitted by an object is equal to the radiant power absorbed by it in thermal equilibrium. Planck's law \cite{ref48} further reveals the basic law of thermal radiation of objects, and quantitatively describes the relationship between blackbody spectral emission and spectral wavelength.
\begin{equation}
\begin{array}{l}
M_{\lambda bb}=\frac{c_1}{\lambda^5}\cdot\frac1{e^{c_2/\lambda T}-1}\label{eqn1}
\end{array}
\end{equation}
where $M_{\lambda bb}$, $\lambda$, $c_1$, $c_2$ and $T$ are the spectral radiant emittance of a blackbody, wavelength, first radiation constant, second radiation constant and thermodynamic temperature, respectively. $c_1=3.7418\times10^8W\cdot\mu m^4/m^2$ and $c_2=1.4388\times10^4\mu m\cdot K$.

Based on Planck’s law, Wien’s law \cite{ref49} describes the relationship between the peak of the spectral emission of the blackbody and the corresponding peak wavelength and the thermodynamic temperature of the blackbody.
\begin{equation}
b=\lambda_mT\label{eqn2}
\end{equation}
where $b$ is a constant and $b=2897.756\mu m\cdot K$.

Stefan-Boltzmann law \cite{ref50} quantitatively describes the relationship between the total radiation exitance of the blackbody and the thermodynamic temperature of the measured object, which can be expressed as Eqn.(\ref{eqn3}).
\begin{equation}
M_{bb}=\sigma T^4\label{eqn3}
\end{equation}
where $\sigma$ is Stefan-Boltzmann constant and $\sigma=5.6704\times10^{-8}W/\left(m^2\cdot K^4\right)$.

\subsection{Principle of Infrared Thermography}
\begin{figure}[!t]
\centering
\includegraphics[width=3.5in]{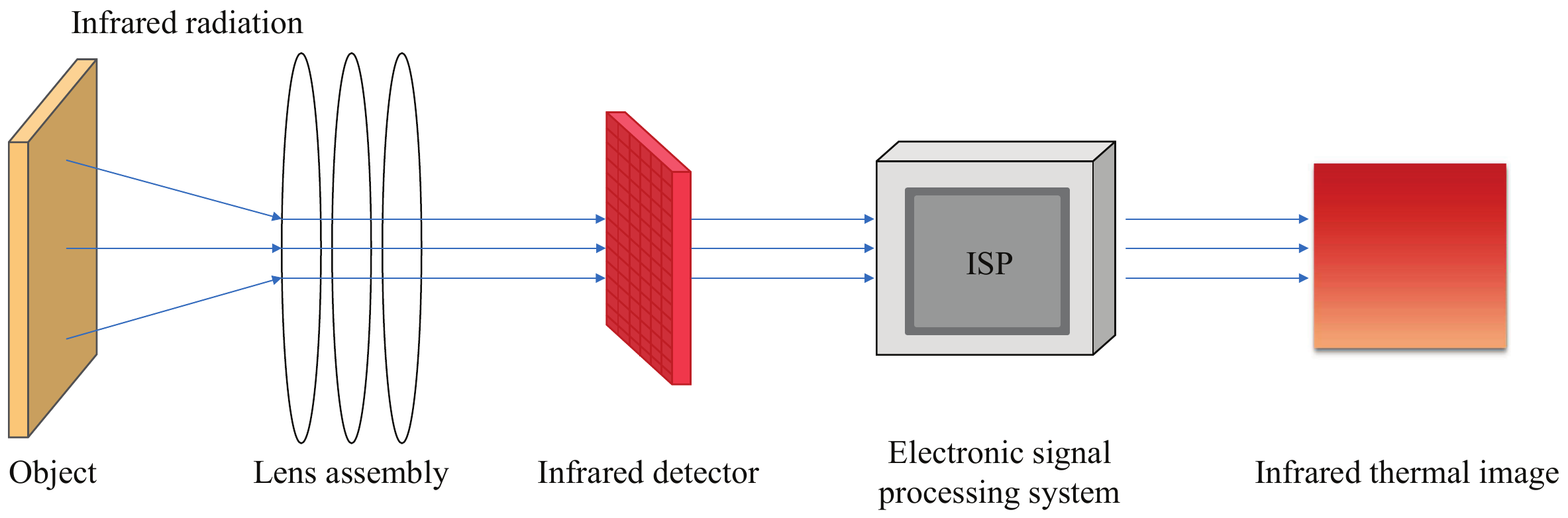}
\caption{Schematic diagram of infrared thermal imaging.}
\label{fig1}
\end{figure}

Infrared thermal imager is a temperature measurement equipment developed according to the principle of infrared radiation. Commonly, an infrared thermal imager comprises four key components: optical imaging lens assembly, infrared detector, electronic signal processing system, and display system, as shown in Fig. \ref{fig1}\cite{ref51}. The optical imaging lens assembly focuses the infrared radiation emitted by the object onto the detector, which receives the radiation and converts it into electrical signals. These signals are then processed by the electronic signal processing system, which translates them into precise temperature information based on predefined calibration algorithms. Finally, the processed temperature information is rendered as a thermal image on the display system, where different colors represent variations in the object's surface temperature. 

\begin{figure}[!t]
\centering
\includegraphics[width=3in]{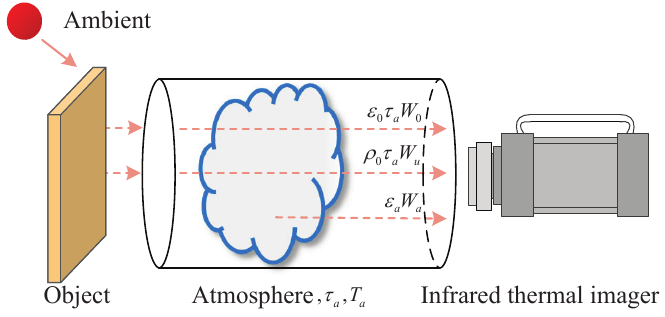}
\caption{Infrared radiation received by the infrared thermal imager.}
\label{fig2}
\end{figure}

Generally, the infrared radiation received by the infrared detector can be divided into three parts: the radiation from the measured object, the radiation from the surroundings and the radiation from the atmosphere, as shown in Fig. \ref{fig2}. The basic temperature measurement principle can be expressed as below \cite{ref52}.
\begin{equation}
W_{rd}\left(T_{r}\right)=\varepsilon_{0}\tau_{a}W_{0}\left(T_{0}\right)+\rho_{0}\tau_{a}W_{u}\left(T_{u}\right)+\varepsilon_{a}W_{a}\left(T_{a}\right)\label{eqn4}
\end{equation}

In (\ref{eqn4}), $W_{rd}$ denotes the infrared radiation received by infrared detector, $W_0$ denotes the infrared radiation emitted by a blackbody at the same temperature as the measured object, $W_u$ denotes the infrared radiation emitted by a blackbody at the same temperature as the surroundings, $W_a$ denotes the infrared radiation emitted by a blackbody with the same temperature as the atmosphere. $\varepsilon_0$ and $\varepsilon_a$ are the emissivity of the measured object and the atmosphere respectively. $\tau_a$ is the atmospheric transmittance. $\rho_0$ is the spectral reflectance of the measured object. $T_r$, $T_0$, $T_u$ and $T_a$ are the radiation temperature, the measured temperature, the ambient temperature and the atmospheric temperature respectively.

According to Kirchhoff’s law, the emissivity of an object in thermal equilibrium is equal to its absorptivity \cite{ref47}. Moreover, the transmittance of an opaque object can be considered to be 0. Therefore, the reflectance of the opaque object can be expressed as Eqn.(\ref{eqn5}). Similarly, the spectral reflectance of the atmosphere can be considered to be 0, and its emissivity can be expressed as Eqn.(\ref{eqn6}).  
\begin{equation}
\rho_0=1-\varepsilon_0\label{eqn5}
\end{equation}
\begin{equation}
\varepsilon_a=1-\tau_a\label{eqn6}
\end{equation}
where $\varepsilon_0$ and $\rho_0$ are the emissivity and reflectance of the measured object, $\varepsilon_a$ and $\tau_a$ are the emissivity and transmittance of the atmosphere.

Then, the infrared radiation received by infrared detector can be expressed as Eqn.(\ref{eqn7}).
\begin{equation}
\begin{aligned}
W_{rd}\left(T_{r}\right)=&\varepsilon_{0}\tau_{a}W_{0}\left(T_{0}\right)+\tau_{a}\left(1-\varepsilon_{0}\right)W_{u}\left(T_{u}\right) \\
& +\left(1-\tau_{a}\right)W_{a}\left(T_{a}\right)\label{eqn7}
\end{aligned}
\end{equation}

According to (\ref{eqn7}), the emissivity of the measured object, ambient temperature, atmospheric temperature and the atmospheric transmittance may cause errors in ITM result. For more detailed information on the principle of IRT, we can refer to the classic work on radiometric temperature measurement \cite{ref53, ref54}.

\section{Analysis of Factors Affecting the ITM Accuracy}
According to the sources of interference factors, this article divides the factors that affect the IRT accuracy into three categories: factors from the external environment, factors from the measured object, and factors from the infrared thermal imager itself. External environmental factors mainly include dust, water mist, ambient temperature, ambient relative humidity, measuring distance, etc \cite{ref55, ref56}. The factors of the measured object mainly refer to the emissivity, source size effect, etc \cite{ref57}. The factors of the infrared equipment itself mainly include the radiation of the infrared detector itself, the temperature drift of the circuit inside the infrared instrument, the field of view of the infrared lens, etc\cite{ref58, ref59}. Table \ref{table1} summarizes the interference mechanisms of these three types of factors and some of the scenarios in which they occur.

\begin{table*}[!t]
\centering
\caption{Common Interference Factors of Infrared Temperature Measurement}
\label{table1}
\renewcommand{\arraystretch}{1.5} 
\begin{tabular}{>{\centering\arraybackslash}m{2.4cm} >{\centering\arraybackslash}m{3cm} m{5.6cm} m{5cm}}
\toprule
\multicolumn{1}{c}{\textbf{Category}} & 
\multicolumn{1}{c}{\textbf{Factors}} & 
\multicolumn{1}{c}{\textbf{Interference Mechanism}} & 
\multicolumn{1}{c}{\textbf{Scenarios with Interference Factors}} \\
\midrule
\multirow{7}{=}{\centering\arraybackslash Factors from the external environment} 
& Dust, water mist, and other atmospheric medium 
& On the one hand, this type of medium may be able to transmit, reflect, and emit radiation, and on the other hand, it may also affect parameters such as atmospheric transmittance 
& Scenarios where there are dust, water mist and other medium interference, such as molten iron temperature measurement in blast furnace ironmaking process \cite{ref60} and steam turbine rotor detection\cite{ref61} \\
\cmidrule(lr){2-4}
& Ambient temperature 
& The object being measured will reflect radiation from the environment, thus affecting the infrared radiation received by the infrared detector 
& Scenarios that is different from the ambient temperature at the time of calibration, such as blast furnace ironmaking process\cite{ref60} \\
\cmidrule(lr){2-4}
& Measuring distance 
& Affects the atmospheric transmittance, the proportion of the measured object in the field of view, etc. 
& Rotary kiln surface temperature measurement and other scenarios where the measuring distance changes \\
\midrule
\multirow{3}{=}{\centering\arraybackslash Factors from the measured object} 
& Emissivity 
& Infrared thermal imagers often set an emissivity, the surface emissivity of different parts of the object being measured may be different 
& Scenarios where the surface emissivity of the measured object is different\cite{ref63} \\
\cmidrule(lr){2-4}
& Source-size-effect 
& The proportion of the measured object in the field of view is different 
& Scenarios where the measured object does not fill the entire field of view \\
\midrule
\multirow{1}{=}{\centering\arraybackslash Factors from the thermal imager itself} 
& Self-radiation of infrared detector, non-uniform response of infrared detector, and so on 
& Influence the conversion calculation of radiation signals 
& Scenarios that focus on detector performance \\
\bottomrule
\end{tabular}
\end{table*}

\subsection{Interference Factors From the External Environment}
Considering that there are many external interference factors in the application of IRT, we did not introduce all the external interference factors one by one, but only listed and analyzed the common atmospheric medium, ambient temperature, measuring distance and other interference factors.
\subsubsection{Atmospheric Medium Interference Factors}
When using IRT under complex and changeable atmospheric conditions, it is necessary to specifically analyze the reflection, scattering, and absorption laws of the particle system in the atmosphere, and then propose corresponding compensation methods to compensate for the ITM results and reduce the measurement errors caused by atmospheric conditions. For example, in some industrial production processes, there are often interference factors such as dust, water vapor, and carbon dioxide due to the characteristics of the production process or the limitations of the installation environment. This section mainly introduces the influence of factors such as dust, water vapor, and carbon dioxide.

\emph{(a) Dust} 

Dust is a common interference factor in industrial processes, which poses a huge challenge to the application of IRT \cite{ref64}. For example, when using IRT to measure the cross-sectional temperature of sintered ore in the sintering machine of the sintering process, the dust stirred up when the sintered ore falls has a significant impact on the ITM results\cite{ref28}.

The dust distributed in the atmosphere can be regarded as a particle system, which could absorb and scatter the infrared radiation passing through it, causing the infrared radiation to attenuate during the propagation process. The dust transmittance can be used to characterize the radiation attenuation caused by dust absorption and scattering. The radiation parameters of the particle system are related to the optical constants (complex refractive index) of the particles contained, the particle content and the particle size distribution \cite{ref65}. In theory, the transmittance, emissivity and reflectance of the dust can be calculated by analyzing the optical constants, the particle content and the particle size distribution. However, the dust conditions in the industrial production process are complex and changeable when using IRT, and it is difficult to obtain these particle parameters in real time and quantitatively, which makes it difficult to calculate the radiation parameters of the particle system.

\begin{figure}[!t]
\centering
\includegraphics[width=3.5in]{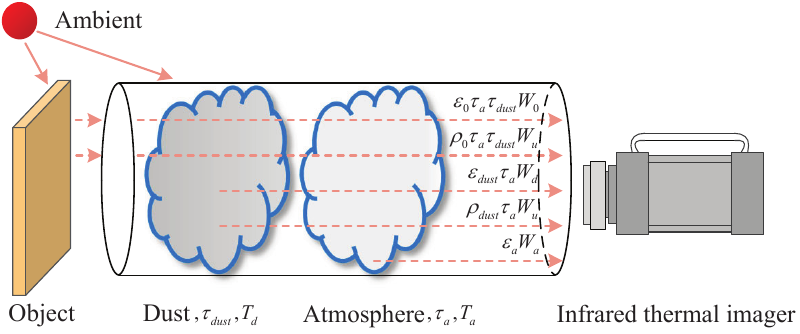}
\caption{Infrared radiation received by the infrared thermal imager when there is dust in the optical path.}
\label{fig3}
\end{figure}

When dust with a certain temperature and concentration is distributed in the optical path between the infrared instrument and the measured object, the radiation received by the infrared detector will change due to the emission and scattering of the dust, thus affecting the final temperature measurement result, as shown in Fig. \ref{fig3}. When there is dust in the optical path, the infrared radiation received by the infrared detector includes five parts: the radiation from the object to be measured, the radiation from the environment reflected by the object to be measured, the radiation from the dust, the radiation from the environment emitted by the dust, and the radiation from the atmosphere \cite{ref66}, as expressed in Eqn.(\ref{eqn8}).
\begin{equation}
\begin{aligned}
W_{rd} = &\varepsilon_{0} \tau_{a} \tau_{dust} W_{0} + \rho_{0} \tau_{a} \tau_{dust} W_{u} + \varepsilon_{dust} \tau_{a} W_{d} \\
& + \rho_{dust} \tau_{a} W_{u} + \varepsilon_{a} W_{a}\label{eqn8}
\end{aligned}
\end{equation}
where $\tau_{dust}$, $\varepsilon_{dust}$, and $\rho_{dust}$ represent the transmittance, emissivity, and reflectance of dust. 

Compared with the radiation received by the infrared detector when there is no dust in the optical path, the presence of dust adds two more infrared radiation items, namely the infrared radiation from the dust and the infrared radiation reflected by the dust. In addition, the dust will change the atmospheric transmittance and cause changes in the original received radiation items. Obviously, the results calculated based on the temperature measurement calibration model embedded in the infrared thermal imager will have large errors.

\emph{(b) Water vapor, carbon dioxide and other gas molecules} 

The atmosphere contains a variety of gas components. According to molecular physics theory, absorption is the result of the interaction between incident radiation and molecular systems\cite{ref67}. The changes in electric dipole moments caused by the vibration of asymmetric molecules such as water vapor, carbon dioxide, ozone, methane, which are present in small amounts in the normal atmosphere, can produce strong infrared absorption. Due to the low ozone concentration in the lower atmosphere, within the wavelength range of more than 1$\mu$m and the altitude range of 12km, the most influential is the selective absorption of infrared radiation by water vapor and carbon dioxide molecules. For example, carbon dioxide has strong absorption bands at 2.7$\mu$m, 4.3$\mu$m, and 15$\mu$m. Other absorbing gases, such as methane, carbon monoxide, nitric oxide, ammonia, hydrogen sulfide, and sulfur oxide, are generally ignored due to their extremely small content in the air. However, it should be pointed out that in some special measurement scenarios, such as blast furnaces, combustion boilers, rotary kilns and other industrial furnaces, it is necessary to analyze the main gas components inside these furnaces according to specific production processes, so as to determine which gas components have a greater impact on infrared radiation and which gas components can be ignored.

The attenuation of infrared radiation caused by atmospheric absorption and scattering can be characterized by transmittance \cite{ref68}. The atmospheric spectral transmittance can be expressed as follows.
\begin{equation}
\tau_a(\lambda)=\tau_1(\lambda)\tau_2(\lambda)\tau_3(\lambda)\label{eqn9}
\end{equation}
where $\tau_1(\lambda)$, $\tau_2(\lambda)$ and $\tau_3(\lambda)$ are the atmospheric spectral transmittance which is absorbed, scattered and attenuated by meteorological conditions respectively.

When the atmospheric environment in the temperature measurement scene is different from the environment in which the infrared thermal imager is calibrated, such as when there are gas molecules such as water vapor and methane, the atmospheric transmittance will change, thereby affecting the infrared temperature measurement results.

\subsubsection{Ambient Temperature}
According to Fig. \ref{fig2}, the infrared radiation received by the infrared detector includes the radiation reflected from the surrounding environment by the measured object. Therefore, when calculating the surface temperature of the measured object, the influence of the ambient temperature must be considered \cite{ref69}. 

According to the change of ambient temperature, this paper divides the ambient temperature into two situations: stable ambient temperature and changing ambient temperature. When the ambient temperature is stable, the error caused by ambient temperature to the temperature measurement result could be reduced by accurately measuring the ambient temperature and considering the influence of ambient temperature in temperature compensation modeling.

When the ambient temperature changes, dynamic compensation methods need to be considered to deal with the impact of ambient temperature changes. When the ambient temperature changes slightly or the ambient temperature is much smaller than the temperature of the measured object, the impact of ambient temperature changes can be ignored while meeting the measurement accuracy requirements. When the ambient temperature changes significantly and its impact cannot be ignored, the ambient temperature change law must first be accurately obtained, and then the measurement results must be compensated according to the ambient temperature to achieve accurate ITM.

It is noted that when the ambient temperature is high and exceeds the working temperature range of the infrared instrument, it is usually necessary to use active cooling methods to cool the instrument, such as cooling nitrogen, circulating water, refrigerators, etc. and make the instrument work within the appropriate temperature range. Otherwise, new temperature measurement errors may be caused due to the influence of the working environment temperature on the detector and processing circuit.

\subsubsection{Measuring Distance}
Infrared thermal imagers can work at a fixed focal length and fixed temperature measurement distance (generally the calibration distance in the laboratory), or in variable zoom and variable distance mode, that is, the focal length changes with the change of measuring distance. For infrared instruments working in fixed focus and fixed distance mode, we can customize the infrared instrument according to the specific measurement requirements, so that the measuring distance of the infrared instrument in actual application is consistent with its calibration distance in the laboratory.

In some practical applications, non-customized infrared instruments are more often used due to the influence of comprehensive factors such as cost and operating requirements. This means that the calibration distance of the infrared instrument is not necessarily consistent with its actual working measuring distance. In this case, it is necessary to consider the impact of the measuring distance on the ITM results.

The influence of measuring distance can be explained from the following infrared temperature measurement principle \cite{ref70}. When the infrared thermal imager is used to measure the surface temperature of an object, the output $s(T_{ob})$ can be represented as follows.
\begin{equation}
s\left(T_{ob}\right)=\pi R_{\max}A_{d}c_{1}\varepsilon_{0}\sin^{2}u_{m}^{'}\int_{\lambda_{1}}^{\lambda_{2}}\frac{\tau_{0}\left(\lambda\right)s\left(\lambda\right)}{\lambda^{5}\left(e^{\frac{c_{2}}{\lambda T_{ob}}}-1\right)}\label{eqn10}
\end{equation}

In (\ref{eqn10}), $A_d$ is the area of a single detection element, $R_{max}$ is the maximum sensitivity of detector, $s(\lambda)$ is the relative sensitivity of the detector, $\lambda_1$ and $\lambda_2$ are working bands of the infrared thermal imager, $c_1$ and $c_2$ are Planck constants. $\tau_o(\lambda)$ is optical system transmittance, $\varepsilon_0$ is the emissivity of the measured object. $\sin u^{'}_m$ is the numerical aperture angle of objective image.

\begin{figure}[!t]
\centering
\includegraphics[width=3.5in]{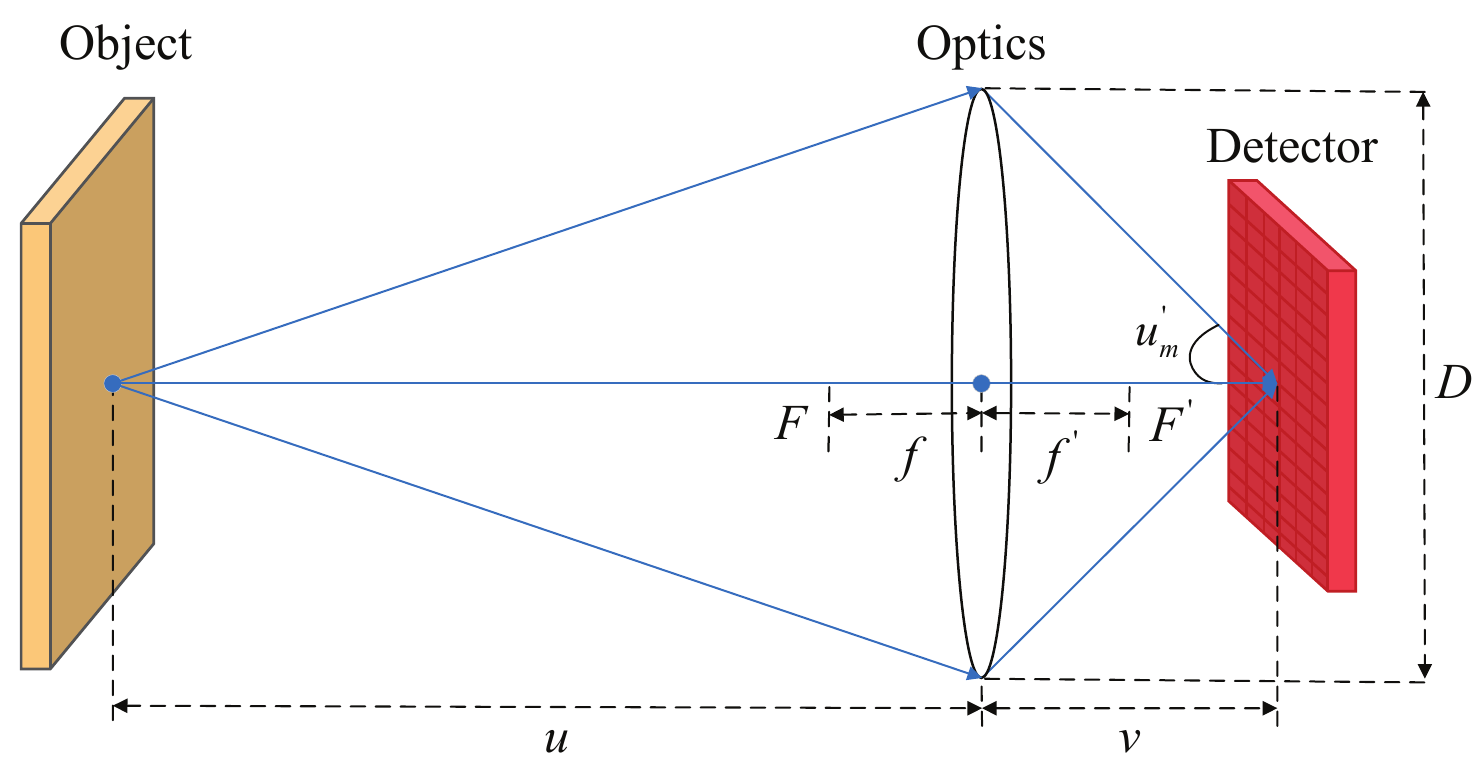}
\caption{Schematic of measured object imaging through optical system.}
\label{fig4}
\end{figure}

In a given infrared system, the detector's area, maximum sensitivity, relative sensitivity, and operating wavelength band are fixed. Therefore, when the surface temperature of the measured object remains constant, any change in the measuring distance primarily affects the output signal of the detector in relation to $\sin u^{'}_m$, as shown in Eqn. (\ref{eqn11}). Fig. \ref{fig4} illustrates the imaging of the measured object through the optical system. Based on the imaging formula, the following equation can be derived.
\begin{equation}
\sin u_m^{'}=\frac{\frac{D}{2}}{\sqrt{\left(\frac{f^2}{u-f^{'}}+f^{'}\right)+\left(\frac{D}{2}\right)^2}}\label{eqn11}
\end{equation}

In (\ref{eqn11}), $D$ represents the pupil diameter of the optical system, $v$ denotes the image distance, $u$ indicates the object distance or the measuring distance, $f$ is the object focus, and $f^{'}$ is the image focus. As shown in (\ref{eqn11}), an increase in measuring distance results in a larger numerical aperture angle of the objective image, which in turn increases the radiation received by the detector plane. Consequently, the measuring distance influences the infrared temperature measurement results. Furthermore, research \cite{ref70} suggests that the measuring distance also affects the proportion of the measured object within the field of view and the atmospheric transmittance, adding complexity to its impact on ITM results.

\subsection{Interference Factors From the Measured Object}
The interference factors from the measured object mainly include the emissivity of the measured object, the source size effect, etc., which are analyzed in detail below.

\subsubsection{Emissivity of the Measured Object}
Emissivity is a key parameter of the object being measured. Emissivity directly affects the object's ability to emit infrared radiation, which in turn affects the infrared radiation entering the infrared detector and corresponding electrical signal, and ultimately affects the ITM results of the object. Therefore, in order to achieve accurate temperature measurement of the object using IRT, it is crucial to obtain the emissivity of the object.

At present, most commercial infrared instruments set a fixed emissivity in advance when measuring temperature. Although this operation meets some temperature measurement needs, it still has a major problem, that is, the emissivity of the measured object may not be fixed. The emissivity of an object is related to its material, shape, surface roughness, degree of oxidation, color, thickness, etc., and the object has different emissivity values at different temperatures and wavelengths. In addition, the emissivity is also related to the test direction, especially when the object being measured has a smooth surface, it is more sensitive to directionality. In order to accurately measure the temperature of the object, when the emissivity of the object being measured is less than 0.9, the infrared instrument is best perpendicular to the surface of the object being measured, and the measurement direction should at least be kept within 30° of the normal direction of the surface of the object, and should not exceed 45° under any circumstances, otherwise the emissivity should be further corrected \cite{ref71}. In other words, the emissivity of an object is a complex function of factors such as material and shape. When using an infrared instrument to measure temperature, simply setting a fixed emissivity may cause non-negligible errors in the measurement results.

When the emissivity of an object can be approximately considered to be a constant, in order to effectively eliminate the measurement error caused by the error in the measurement background and the actual temperature of the target, the double reference method, double temperature method, double background method and other methods can be considered to obtain the emissivity of the measured object\cite{ref72}.

In addition, there is a more practical method in industry \cite{ref73}. First of all, the temperature of the measured object is accurately obtained by using other reference sensors, such as thermocouples. Then, the infrared instrument is used in the same position to measure the same area of the measured object, and the set emissivity of the infrared instrument is adjusted until it is adjusted to a certain emissivity so that the measurement result of the infrared instrument is consistent with the measurement result of other reference sensors. At this time, the set emissivity can be considered as the effective emissivity of the measured object.

When the emissivity of an object is in a changing state, such as when the temperature of the measured object is unstable or the surface state is changing, these situations will undoubtedly increase the difficulty of accurate temperature measurement by infrared instruments. Because most of the commercial infrared instruments can only set a certain emissivity, and cannot dynamically adjust the emissivity according to the change of the emissivity of the object. In this case, it is necessary to consider the temperature compensation methods to correct the temperature measurement error caused by the emissivity. When studying the infrared temperature measurement compensation method, it is necessary to first consider the change law of the emissivity of the object.

At present, it can be seen from the literature that there are few reports on infrared temperature measurement methods for changing emissivity. In different production processes, the emissivity changes of different objects are different, which may make it difficult to propose a universal infrared temperature measurement compensation method for changing emissivity.

\subsubsection{Size-of-Source Effect}
All infrared thermal imagers are affected by the Size-of-Source Effect (SSE) to a certain extent during measurement \cite{ref74}. Scattering and diffraction within the optical system of the infrared thermal imager cause a part of the radiation within the nominal field of view to be lost, while another part of the radiation outside the nominal field of view is detected. The effect on the infrared imager signal depends on the size of the target and the radiation distribution around the target. From this point of view, the source-size-effect is attributed to the interference factor from the measured object.

In fact, SSE causes the actual field of view of the infrared thermal camera to be uncertain. SSE becomes a problem when IRT is used to measure different targets. These targets may be objects whose temperature some industrial process needs to obtain, or they may be blackbodies used to calibrate radiation thermometers. Therefore, it is necessary to amend SSE.

Saunders systematically analyzed the source-size-effect, formalized the derivation of the equation used to correct the SSE radiation thermometer signal, and obtained four specific equations for the SSE quantities that are more commonly used. For more information on SSE, we can refer to P. Saunders' study \cite{ref75}.

\subsection{Interference Factors From the Infrared Thermal Imager Itself}
This section mainly introduces the interference factors related to the infrared instrument itself, including the self-radiation of the infrared detector, the non-uniform response of the infrared detector, the temperature drift of the circuit inside the infrared instruments, the angle of view of the infrared instruments, and the field of view of the infrared lens.

In theory, the infrared thermal imager has been calibrated before leaving the factory, and the infrared thermal imager itself will not affect the infrared temperature measurement results. However, if the working time of the infrared thermal imager is too long or the ambient temperature is high, for the uncooled infrared thermal imager, the infrared detector's own radiation and the circuit temperature drift inside the instrument will affect the conversion calculation of the radiation signal, and ultimately affect the temperature measurement results. Therefore, for the temperature measurement scenario with high ambient temperature, it is necessary to cool the infrared thermal imagers and avoid the influence of circuit temperature drift on the ITM results.

In addition, some scholars have pointed out the influence of the angle of view on the ITM results. From the temperature measurement results, it can be concluded that the angle of view is different and the measurement results are different\cite{ref76}. In essence, the angle of view affects the directional emissivity. Therefore, from this point of view, the effect of the angle of view can be regarded as the effect of the directional emissivity. Pablo \cite{ref77} pointed out that the field of view of the infrared thermal imager affected the proportion of the measured object in the field of view and further influenced the ITM results. In this sense, the effect of the field of view is similar to that of the SSE.

\section{Infrared Temperature Measurement Compensation Method}
In view of the influence of interference factors on ITM, many scholars have carried out research on ITM compensation methods, which has enriched the theory of infrared temperature measurement. From the angle of compensation means, these compensation methods can be classified into hardware compensation \cite{ref78} and software compensation \cite{ref79}. Hardware compensation mainly improves the detector performance from the hardware point of view, while software compensation realizes temperature compensation by software algorithms. Since hardware compensation cannot eliminate the influence of external environment interference factors, more research has been done on software compensation methods in recent years. Generally, software compensation mainly builds a temperature compensation model to eliminate or reduce the ITM error caused by interference factors based on infrared temperature measurement mechanism or temperature measurement data and does not need to add additional hardware. This paper mainly discusses the research of software compensation. The infrared thermometer can be regarded as the infrared thermal imager with smaller detector array, so the review includes some research work using the infrared thermometer.

This work divides the software compensation methods into three categories according to their compensation ideas: 1) Mechanism Modeling-based Compensation method (MMC), 2) Data-Driven Compensation method (DDC), 3) Mechanism and Data jointly driven Compensation method (MDC). This section focuses on the investigation of software compensation methods, and summarizes the evaluation indicators of compensation methods. Table \ref{table2} summarizes the compensation principles, advantages and disadvantages of different compensation methods and their corresponding application scenarios.

\begin{table*}[!t]
\centering
\caption{Comparison of Different Compensation Methods}
\label{table2}
\renewcommand{\arraystretch}{1.5} 
\begin{tabular}{m{0.7cm} m{3.3cm} m{4cm} m{4.2cm} m{3.8cm}}
\toprule
\multicolumn{1}{c}{\textbf{Types}} & \multicolumn{1}{c}{\textbf{Compensation Principle}} & \multicolumn{1}{c}{\textbf{Advantages}} & \multicolumn{1}{c}{\textbf{Disadvantages}} & \multicolumn{1}{c}{\textbf{Application Scenarios}} \\
\midrule
MMC & Analyze the influence of interference factors on infrared radiation, and establish the compensation model based on infrared temperature measurement mechanism 
& Theoretically, the temperature measurement error caused by interference factors can be fundamentally eliminated, and the interpretation is strong 
& The mechanism is complex, the model is difficult to establish, and the parameters in the model are difficult to obtain accurately 
& Infrared temperature measurement scenarios where the interference factor is single and easy to quantify, and the interference mechanism is relatively clear \\
\midrule
DDC & Establish the nonlinear mapping relationship between the parameters related to interference factors and the desired temperature information 
& The modeling process is simple and does not require understanding of complex infrared temperature measurement principles 
& The interpretability of the model is weak, the model performance depends on the quality and scale of the data set, and sometimes the data set is difficult to obtain 
& Infrared temperature measurement scenarios in which interference factors are easy to quantify and temperature measurement data is easy to collect \\
\midrule
MDC & Both infrared temperature measurement mechanism and data-driven modeling method are used when constructing the temperature compensation model 
& The advantages of infrared temperature measurement principles and data-driven modeling are combined 
& The combination of infrared temperature measurement principles and data-driven modeling is difficult and there are less related studies 
& Infrared temperature measurement scenarios in which interference factors are easy to quantify, interference mechanism is clear and temperature measurement data is easy to collect \\
\bottomrule
\end{tabular}
\end{table*}

\subsection{Mechanism Modeling-Based Compensation Method}
The mechanism modeling-based compensation method (MMC) analyzes the interference mechanism of the influencing factors, introduces the radiation terms of the interference factors into the infrared radiation temperature measurement model, and constructs an ITM compensation model for the interference factors to achieve compensation for the ITM results.

In view of the influence of dust on ITM, Pan \emph{et al.}  \cite{ref66} studied a compensation method based on the principle of infrared temperature measurement, introduced dust radiation terms when constructing the radiation temperature measurement model, and proposed a dust parameter estimation method based on a reference body. The effectiveness of the compensation method was demonstrated through two experiments in laboratory. It should be pointed out that this method has not been applied in industry, and its applicability in real dust scenes needs further exploration.

Reference \cite{ref80} considered the influence of ambient temperature on ITM results, analyzed the main uncertainties of the system, type A (evaluated with statistical methods) and type B (nonstatistical), fitted the voltage change caused by ambient temperature with a ninth-order polynomial, and established a compensation function for ambient temperature based on the ITM principle, which improved the measurement accuracy of the infrared thermometer within the temperature measurement range of -40-60$^{\circ}$C. Shu \emph{et al.} \cite{ref81} analyzed the influence of ambient temperature on the temperature measurement of rotary kiln, and found that ambient temperature mainly affects the reflected radiation and atmospheric radiation of the environment, resulting in a decrease in temperature measurement accuracy. Then, a dynamic temperature compensation model was proposed based on the uneven change of temperature field and  the temperature measurement accuracy was improved by 5$\%$. 

Zhang \emph{et al.}  \cite{ref82} analyzed the impact of atmospheric transmittance on the measurement accuracy of infrared thermal imagers during long-distance temperature measurement, quantified the attenuation effects of water vapor, carbon dioxide, aerosols, rain and snow and other weather conditions, and established an atmospheric transmittance model for infrared thermal imagers and a temperature measurement model at sea level. They also demonstrated through experiments that this method reduced the influence of atmospheric transmittance and improved the measurement accuracy of infrared thermal imagers. 

Wang \emph{et al.} \cite{ref83} applied infrared radiation temperature measurement technology to the measurement of explosion fireball temperature and proposed a temperature compensation model based on geometric optics and infrared radiation theory. This model eliminated the measurement error of the test environment on the radiation energy, analyzed the influence of distance and meteorological conditions on the measurement accuracy, introduced image grayscale as an intermediate variable, and derived the temperature compensation relationship based on the error transfer theory.

In response to the influence of emissivity, Pan \emph{et al.} \cite{ref84} constructed an ITM compensation model for directional emissivity changes based on the ITM mechanism, and proposed a directional emissivity correction method based on a reference body using a three-dimensional thermal imaging system constructed by an infrared thermal imager and a laser scanner, which effectively reduced the ITM error caused by directional emissivity changes. Shen \emph{et al.} \cite{ref85} studied the influence of the surface emissivity of an object on the ITM accuracy, further analyzed the influence of materials, surface conditions, and temperature on emissivity, and established an emissivity model from the perspective of the infrared thermal imager. On this basis, a non-steady-state temperature field model of infrared thermal images with different emissivity was established, and the emissivity corresponding to each point was calculated, eliminating the influence of temperature and surface conditions on emissivity, thereby reducing the influence of emissivity on ITM accuracy. Furthermore, cast iron and stainless steel were used as the measured objects to verify the feasibility of their method.

Saunders \cite{ref75} analyzed the influence of the source-size-effect (SSE), introduced simple and complex SSE general correction methods, and gave a method for determining SSE quantities. Dai \emph{et al.}  \cite{ref86} designed a high-precision infrared temperature measurement algorithm based on the ITM principle to solve the problem of nonlinear response of pixels in the infrared focal plane, which significantly reduced the ITM error. 

From the above work, it can be seen that some scholars have carried out research on the MMC methods and achieved certain application effects. The MMC method can fundamentally eliminate the influence of interference factors on ITM results, ensure the reliability of the results, and have clear interpretability. In some temperature measurement scenarios where the interference factor is single and easy to quantify, the MMC method is a good choice, and a specific mechanism model can be established to eliminate the influence of this factor. However, it is noted that most of the aforementioned MMC methods are used to compensate for the influence of a single interference factor. For complex scenarios with multiple interference factors, there are few reports on the MMC method. 

In addition, the mechanism model itself is often difficult to construct or obtain, especially when using commercial infrared thermal imagers, whose output is temperature information. Typically, the specific radiation temperature measurement model is integrated within the infrared thermal imager and remains inaccessible to users directly, complicating the establishment of an accurate compensation model based on the ITM mechanism.

Moreover, another challenge that limits the use of MMC is that the parameters in the mechanism model are difficult to obtain or quantify, such as dust transmittance, dust reflectivity, emissivity of the measured object, and ambient temperature. The acquisition of these parameters depends on more detailed parameters such as the optical constants of the dust particle system, particle size distribution, temperature and state of the object being measured, etc. These parameters may be difficult to obtain in the actual production process, or can only be looked up in a table or estimated based on experience. The performance of MMC methods depends on the accuracy of the above parameters. If these parameters have large errors, it is difficult to guarantee the effect of the final compensation method, and may even make the compensated ITM results unusable. Consequently, in temperature measurement scenarios involving complex interference factors that make it challenging to construct a mechanism model, MMC may not be the most suitable choice for temperature measurement.

\subsection{Data-Driven Compensation Method}
Data-driven modeling has been a hot topic in recent years \cite{ref87, ref88, ref89, ref90, ref91}. Especially with the development of artificial intelligence and computing power, many scholars have also carried out research on data-driven infrared temperature measurement compensation methods \cite{ref92, ref93, ref94, ref95, ref96, ref97, ref98, ref99, ref100, ref101, ref102}. The DDC method quantifies the interference factors and directly constructs a mapping model between infrared temperature measurement errors and interference factors based on data-driven models such as polynomial fitting and neural networks, thereby achieving compensation for ITM results.

Bao \emph{et al.} \cite{ref92} proposed a heat-assisted detection and ranging (HADAR) method, and constructed a TeX-Net with hyperspectral heat cube as input, which predicted the surface temperature, emissivity, texture and other information of the measured object, and achieved good infrared temperature measurement accuracy. In view of the influence of ambient temperature, Shen \emph{et al.} \cite{ref93} constructed an infrared radiation measurement module containing multiple temperature sensors, detected the ambient temperature through a temperature sensor, and constructed a compensation model by using polynomial fitting.

Pan \emph{et al.} \cite{ref94} studied the influence of multiple interference factors on infrared temperature measurement, taking measuring distance and dust as interference factors, and constructed a compensation model for ITM errors based on nonlinear polynomial regression. The effectiveness of the proposed method was demonstrated through laboratory experiments. Furthermore, Pan \emph{et al.} \cite{ref95} introduced artificial intelligent into temperature compensation, constructed an intelligent compensation model based on weighted ensemble stacked denoising autoencoder to address the influence of measuring distance and dust on ITM, and verified the effectiveness of the method through blackbody furnace and metal cup experiments.

Regarding the influence of measuring distance, ambient temperature and visibility on the ITM accuracy, Liang \emph{et al.} \cite{ref96} compared the effects of polynomial fitting and neural network fitting. The experiment showed that the accuracy of neural network fitting is higher than that of traditional polynomial fitting. Zhang \emph{et al.} \cite{ref97} considered the influence of detector non-uniformity and ambient temperature, established the relationship between infrared image pixel value and blackbody temperature, ambient temperature and thermal imager temperature, and carried out experiments with blackbody furnace as the measured object, showing that their method can accurately measure the absolute temperature of the blackbody furnace target surface.

Aiming at the influence of measuring distance, Zhang \emph{et al.} \cite{ref70} explained the influence of measuring distance on ITM accuracy from three perspectives: field of view angle, temperature difference between background and temperature measurement object, and atmospheric transmittance. Then, a data-driven compensation model for these three factors was constructed, and the average of compensation model was used as a compensation model for temperature measurement distance. The effectiveness of their method in reducing the influence of measuring distance is verified in a blackbody furnace experiment. In \cite{ref98}, in order to meet the need of real-time detection of human body surface temperature, a non-contact real-time temperature measurement method based on distance compensation is proposed. The distance data is collected by laser ranging module, and the compensation formula of temperature with distance is determined according to the least square algorithm. When performing ITM on a rotary kiln, the measuring distance will affect the ITM results. Guo \emph{et al.} \cite{ref99} used a cubic polynomial to fit the temperature difference data, and used the obtained function to compensate for the temperature with error, which significantly improved the ITM accuracy. Reference \cite{ref100} designed and completed a comparative experiment on the temperature measurement of a blackbody furnace at different measuring distances, different measuring angles, and different working times. Polynomial fitting was used to characterize the influence of measuring distance, measuring angle, and additional thermal radiation of the thermal imager on the ITM accuracy, thereby improving the ITM accuracy of the thermal imager at close distances.

Reference \cite{ref77} demonstrated through ITM experiments that the field of view of the infrared thermal imager lens and the different viewing angles between the analyzed object and the imager will introduce errors into the ITM results. On this basis, an experimental model based on artificial neural network and regression was established to correct these errors. In view of the influence of viewing angle on the measurement of animal surface temperature, Jiao \emph{et al.} \cite{ref101} used an infrared thermal imager and Kinect sensor to obtain the infrared thermal image and depth information of the animal, and then established the relationship between the reference temperature and the measured temperature and the measurement angle based on multivariate regression, which significantly reduced the temperature measurement error caused by the viewing angle. Aiming at the influence of the incident angle on the ITM results, an incident angle compensation algorithm based on polynomial fitting was established to quantify the functional relationship between the true temperature and the temperature with error at different incident angles \cite{ref102}.

From the above work, it can be seen that many scholars have conducted research on DDC methods, which provides good support for ensuring the ITM accuracy. The DDC method does not require understanding of the complex infrared temperature measurement principle. Instead, it directly establishes the relationship between the parameters related to the interference factors and the expected information through nonlinear fitting methods such as polynomial regression and neural network, and achieves good results in many scenarios. In infrared temperature measurement scenarios where interference factors are easy to quantify and the temperature measurement data is relatively easy to collect, the DDC method is a good choice. However, it should be pointed out that the DDC model mainly determines the structure or parameters of the model based on the final compensation effect, and the interpretability is weak. As a result, it is often difficult to use the DDC model to explore the interference mechanism of interference factors on ITM results. 

Furthermore, the performance of DDC methods depends on the quality and scale of the data set used. If the quality of the data set is high, the compensation model may be effective. On the contrary, if the quality of the data set is very poor and contains a lot of noise, the compensation model obtained by training may not be effective. Data scale is also a problem. For example, the compensation model constructed from temperature measurement data collected in the 40-120℃ temperature range may not be applicable to the application scenario of the 120-300℃ temperature range. What is more, the difficulty of collecting and constructing the data set is also a factor to consider. Data collection in a laboratory environment may be relatively easy, while in many industrial scenarios, such as the blast furnace ironmaking process, data collection is a very challenging task.

\subsection{Mechanism and Data Jointly Driven Compensation Method}
Through the above analysis of the MMC and DDC methods, it can be seen that the first compensation method based on the infrared temperature measurement principle is more explanatory and universal, but it faces the dilemma that some parameters are difficult to determine or the establishment of MMC is full of challenges. The second method does not require detailed understanding of the infrared temperature measurement principle, and uses the test data related to interference factors to achieve compensation for ITM errors. However, DDC lacks analysis on the principle of infrared temperature measurement and has poor universality, so different test data should be obtained in different scenarios to build compensation models. Therefore, some scholars have explored the mechanism and data jointly driven compensation (MDC) methods.

According to the available literature, there are few MDC methods. Li \emph{et al.}  \cite{ref103} used the rich texture features of visible images to build a data-driven calculation model of water mist transmittance, and combined with the infrared temperature measurement mechanism model, realized accurate ITM under dynamic water mist interference. Zhang \emph{et al.}  \cite{ref81} studied the influence of atmospheric transmittance on the ITM accuracy, calculated the relationship between atmospheric transmittance and water vapor, carbon dioxide, aerosol, etc., by using curve fitting, and then compensated the temperature measurement results according to the infrared temperature measurement mechanism model. These studies provide some inspirations for ITM compensation, but more exploration and research on MDC methods are still needed. For example, for some parameters that are difficult to determine in the MMC methods, a data-driven method can be considered to determine these parameters, so as to enhance the adaptability of the compensation method. For the DDC methods, the analysis of infrared temperature measurement principle can be considered to enhance the interpretability of the compensation method.

\subsection{Evaluation of Compensation Methods}
After compensating the infrared temperature measurement result, it is essential to conduct error analysis of the compensated results to assess the improvement in measurement quality. Additionally, performing sensitivity and uncertainty analysis on the compensation methods is also necessary. Sensitivity analysis evaluates the influence of key parameters on the compensation method, while uncertainty analysis estimates the comprehensive uncertainty of the compensation result due to the uncertainty component in the compensation method.

\subsubsection{Error Analysis}
Error analysis can demonstrate the extent to which compensation methods enhance temperature measurement accuracy by comparing the results obtained before and after compensation. Two commonly used metrics are the mean absolute error (MAE) and the root mean square error (RMSE) \cite{ref104}.
\begin{equation}
\mathrm{MAE}=\frac{1}{N}\sum_{i=1}^{N}\left|T_{i}-T_{i}^{'}\right|
\end{equation}
\begin{equation}
\mathrm{RMSE}=\left[\frac{1}{N}\sum_{i=1}^{N}\Bigl(T_{i}-T_{i}^{'}\Bigr)^{2}\right]^{1/2}
\end{equation}
where $T_i$ and $T_{i}^{'}$ represent the true temperature of the object and the measured temperature, respectively, and $N$ is the number of compensations applied to the original measurement results.

By calculating the MAE and RMSE before and after compensation, the ITM errors can be quantitatively displayed, providing a clear visualization of the improvement in temperature measurement accuracy achieved through the application of compensation methods.

\subsubsection{Sensitivity Analysis}
Sensitivity analysis examines how variations in system parameters or environmental conditions affect the state or output of a system or model. This is achieved by changing a parameter of the model and analyzing the resultant changes in the model's output \cite{ref105}. Typically, not every parameter in the model is subjected to sensitivity analysis. Instead, only those parameters with significant uncertainties are analyzed.

When performing sensitivity analysis on ITM compensation methods, parameters with substantial uncertainties should be analyzed according to the specific situation, such as the dust transmittance or the proportional coefficients between measured and ambient temperature are selected for analysis in \cite{ref66}. By performing sensitivity analysis on a compensation method with respect to a certain parameter, the sensitivity of the compensation method can be learned. If altering a particular parameter result in minimal changes to the compensated temperature, it indicates that the parameter has a minor impact on the model, thereby suggesting high stability of the compensation method and vice versa.

\subsubsection{Uncertainty Analysis}
Measurement uncertainty characterizes the range within which the true value of the measured result is estimated to lie, representing a parameter of the measurement result that indicates its dispersion \cite{ref106}. According to the definition of measurement uncertainty, a complete measurement result comprises both the estimated value and a dispersion parameter. For instance, the measurement result of a quantity $Y$ is expressed as $Y \pm U$, where $Y$ is the estimated value, and $U$ is the measurement uncertainty. Under this definition, the measurement result does not represent a single definite value but rather an interval encompassing infinitely many possible values. Measurement uncertainty is a crucial indicator of the quality of measurement results. lower uncertainty signifies higher quality and greater utility, while higher uncertainty indicates lower quality and utility.

Conducting uncertainty analysis on compensated ITM results allows for the evaluation of the compensation method's effectiveness. Furthermore, in certain processes, the temperature of the measured object must remain within a specific range to prevent accidents such as explosions or mechanical failures. By assessing the uncertainty of the compensated results, we can determine whether the compensation meets practical requirements and whether there is a risk of temperature exceeding safe limits. This is particularly important for production processes with high accuracy requirements where inaccurate temperature measurements could lead to severe consequences. Therefore, uncertainty analysis of compensated temperature is essential.

In practical measurement processes, various factors affect the accuracy of measurement results. Consequently, measurement uncertainty typically comprises several components, and each uncertainty component can be evaluated by two types of methods, Type A evaluation and Type B evaluation, regardless of its nature. Type A evaluation is based on statistical analysis of a series of observations, while Type B evaluation relies on prior information or assumed probability distributions. 

The uncertainty represented by standard deviation is called standard uncertainty. The type A evaluation of the standard uncertainty is assessed using statistical analysis, and the standard uncertainty is equivalent to the standard deviation obtained from the series of observations. Type B evaluation of standard uncertainty does not use statistical analysis, but estimates probability distribution or distribution hypothesis based on other information to evaluate standard deviation and obtain standard uncertainty. Because some uncertainties cannot be evaluated by statistical methods, or although statistical methods can be used, it is not economically feasible, so type B evaluation is mostly used in practical work.

For further theoretical background on uncertainty analysis, Infrared Thermography Errors and Uncertainties \cite{ref107} and Measurement Uncertainty in Error Theory and Data Processing \cite{ref108} are good references.

\section{Main Issues and Prospects}
Despite significant advancements in existing ITM compensation methods, challenges remain when applying IRT in complex scenarios. This section discusses the main issues associated with current compensation methods and outlines some future research directions.
\subsection{Main issues}
\subsubsection{Incomplete Elimination of Measurement Errors Caused by Interference Factors}
Current compensation methods for ITM reduce, to some extent, the errors caused by interference factors but cannot eliminate them. For MMC methods, accurately determining the parameters of the model during compensation modeling is challenging. DDC methods face difficulties in ensuring that the dataset encompasses all possible interference factors. Consequently, it is difficult to say that a compensation method can completely eliminate the influence of interfering factors. However, researchers can employ more advanced and rigorous compensation methods to minimize the measurement errors caused by these factors as much as possible.

\subsubsection{Limited Techniques for Identifying and Quantifying Interference Factors}
Difficulty in identifying and quantifying interference factors is a significant obstacle in constructing accurate ITM compensation models. Existing IRT techniques lack the capability to accurately identify interference factors. In complex environments, environmental factors such as dust and water mist are important factors affecting the ITM accuracy, and these interferences have the characteristics of dynamic non-uniform distribution. Although IRT can capture two-dimensional temperature distributions of the target object, it is highly susceptible to environmental interferences. Thus, it is necessary to accurately identify whether there is environmental interference and then determine whether temperature compensation is required. In addition, the real-time quantification of physical parameters such as concentration and transmittance of environmental interference factors is the basis for targeted compensation of ITM errors. However, infrared thermal imagers do not have the function of quantifying physical parameters of environmental interference factors. How to quantify interference factors on the basis of identifying interference factors is worth thinking about.

\subsubsection{Insufficient Consideration of the Coupling Effects of Multiple Interference Factors}
In real-world applications, there may be multiple interference factors in the temperature measurement scene and the ITM error is the result of the combined effect of multiple interference factors. For example, in the molten iron temperature measurement of blast furnace ironmaking process, the ITM results of molten iron will be affected by dynamic dust and ambient temperature at the same time. However, most of the existing ITM compensation methods address only single interference factor, limiting their application in scenarios where multiple factors are present. In addition, the coupling relationship between multiple interference factors also needs to be considered, which may affect subsequent error analysis and uncertainty analysis. Some studies simplify the combined effects by linearly summing the individual influences of factors. For instance, this work equates the impact of ambient temperature and measuring distance to the sum of the impact of ambient temperature and the impact of measuring distance on ITM results \cite{ref109}. This direct summation mode assumes that the interference factors are independent of each other, but in fact, the impact of multiple interference factors on ITM results may be coupled with each other.

\subsubsection{Underutilization of Multi-Source Information}
In the application of IRT technology, infrared thermal imager is the main temperature measurement instrument. Although infrared thermal imager can obtain the two-dimensional temperature distribution of the measured object, its ability to obtain other information is limited. Many compensation methods utilize only the data obtained from infrared images and neglect other potentially valuable information such as dust transmittance, water mist transmittance, and smoke concentration, which complicates the construction of compensation models and makes model parameter determination challenging. 

\subsubsection{Difficulty in Introducing Artificial Intelligence Algorithms}
At present, artificial intelligence (AI) is a hot topic in many fields, which has brought a lot of changes to the research directions of target detection, image fusion, condition recognition, soft sensor, process monitoring, fault diagnosis and so on. An important basis for the implementation of artificial intelligence algorithms is datasets. However, it is often difficult to establish datasets during ITM compensation modeling. Because in the real temperature measurement scene, many data related to interference factors are difficult to collect, or it is difficult to simulate the temperature measurement scene under the influence of interference factors in high fidelity in the laboratory. In addition, there are few research on the combination of artificial intelligence such as deep learning and ITM compensation modeling. The limitations of data sets and related references make it difficult to introduce artificial intelligence algorithms in compensation modeling, which hinders the development of intelligent compensation methods for ITM.

\subsection{Prospects}
Considering the existing problems of ITM compensation methods, it is necessary to further explore and study the accurate infrared temperature measurement method, and the future research directions are discussed in this section.

\subsubsection{Mechanism and Data Jointly Driven Compensation Modeling}
Existing compensation methods typically rely either on radiative temperature measurement principles or data-driven approaches. Mechanism models often struggle with parameter determination, while data-driven models face challenges related to extensive data collection or data acquisition. Mechanism and data jointly driven compensation models offer a promising direction for ITM compensation modeling. For instance, in the literature \cite{ref44}, the transmittance of water mist is determined based on the data-driven model, and the temperature compensation model is constructed based on the infrared temperature measurement principle. Then, the calculated transmittance of water mist is substituted into the temperature compensation model to reduce the infrared temperature measurement error caused by dynamic water mist. Therefore, we can consider the combination of mechanism modeling and data-driven approaches to compensate for the ITM errors. For example, using the data-driven model to help calculate some parameters required by the mechanism model, and introducing the mechanism model to expand the data set required by the data-driven model.

\subsubsection{Identification and Quantification of Environmental Interference Factors}
Recognizing and quantifying environmental interference factors within the measurement scenario provides essential prior environmental information for IRT. This involves two aspects of research. (a) Identification of environmental interference factors. In complex scenarios with intermittently and dynamically distributed factors such as dust and water mist, it is crucial to develop methods for accurately identifying these factors to enable targeted dynamic compensation. (b) Real-time quantification of environmental interference factors. Many environmental factors exhibit non-uniform distributions, such as dust concentrations. Real-time quantification of parameters like concentration and transmittance is fundamental for dynamic compensation of ITM errors. Especially with the development of AI and object detection algorithms, it is possible to use gas object detection algorithm to identify and quantify atmospheric medium interference factors such as dust.

\subsubsection{Research on Compensation Methods for Multiple Interference Factors}
Most current compensation methods address single interference factors, whereas real-world scenarios often involve multiple simultaneous interferences, such as varying measuring distances and the presence of dust and water mist. Furthermore, the interactions between multiple factors, which are frequently coupled rather than independent, need to be comprehensively considered. Therefore, the temperature compensation model for multiple interference factors needs to be further studied to ensure the ITM accuracy, such as the introduction of deep learning networks to explore the influence of multiple interference factors.

\subsubsection{Multi-Source Information Collaborative Measurement}
Existing compensation methods typically rely on a single source of information, such as infrared thermal imager, which limits the potential for comprehensive compensation. Multi-source information detection can provide complementary data that enhances compensation accuracy. For example, the study by Pan \emph{et al.} \cite{ref44} introduced a novel compensation method that combines infrared and visible light vision and utilized the visible light information as prior environmental data for infrared temperature compensation. Therefore, integrating infrared thermal image data with other relevant information sources, such as dust transmittance and water mist concentration, can exploit the complementary and synergistic relationships between different data types, thereby improving the ITM accuracy under interference.

\subsubsection{Introduction of Artificial Intelligence Algorithms}
AI algorithms have proven to be powerful tools across various domains. In the ITM compensation modeling, AI can leverage its robust nonlinear mapping capabilities to quantify the relationship between interference factors and ITM errors. Regarding the difficulty in constructing the dataset required for compensation modeling, it is also possible to consider using generative intelligent models, simulation software, laboratory simulation of interference factors, etc. to construct the datasets required for training the intelligent compensation model, thereby providing a data basis for the intelligent ITM compensation model.

\section{Conclusion}
Temperature measurement accuracy is an important issue in the IRT application, and it directly affects the subsequent tasks that use the temperature measurement results. This article analyzes common interference factors when using IRT, including external environmental interferences such as atmospheric media, ambient temperature, and measurement distance, object-related interferences like emissivity and intrinsic interferences from the infrared detector itself, such as self-radiation and circuit drift. Furthermore, we summarize compensation methods based on mechanism modeling, data-driven modeling, and mechanism and data jointly driven modeling. Despite the availability of many effective compensation methods, increasingly complex application scenarios and higher accuracy requirements present new challenges for researchers. Therefore, in the field of ITM compensation, continuous exploration and close cooperation of scientific research and engineering personnel are still needed to deeply study the joint mechanism and data compensation modeling, identification and quantification of environmental interference factors, compensation modeling for multiple interference factors, multi-source information collaborative measurement, and introduction of artificial intelligence. Solving these problems will help further reduce the impact of interference factors on ITM, expand the application scenarios of IRT, and provide accurate and reliable infrared temperature measurement solutions for more temperature measurement scenarios.

\ifCLASSOPTIONcaptionsoff
  \newpage
\fi

%

\begin{IEEEbiography}
[{\includegraphics[width=1in,height=1.25in,clip,keepaspectratio]{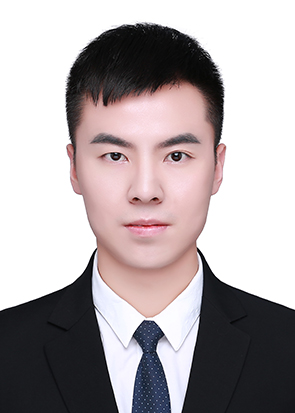}}]{Dong Pan}
received his B.S. degree in Automatic and the Ph. D degree in control science and engineering from Central South University, Changsha, China, in 2015 and 2021, respectively. From 2019 to 2021, he was a visiting scholar with the Department of Electrical and Computing Engineering of Université Laval, Québec City, Canada. 

He is currently an associate professor at Central South University. His main research interests include infrared thermography, temperature measurement, image processing, and modeling and control of industrial processes.
\end{IEEEbiography}

\begin{IEEEbiography}[{\includegraphics[width=1in,height=1.25in,clip,keepaspectratio]{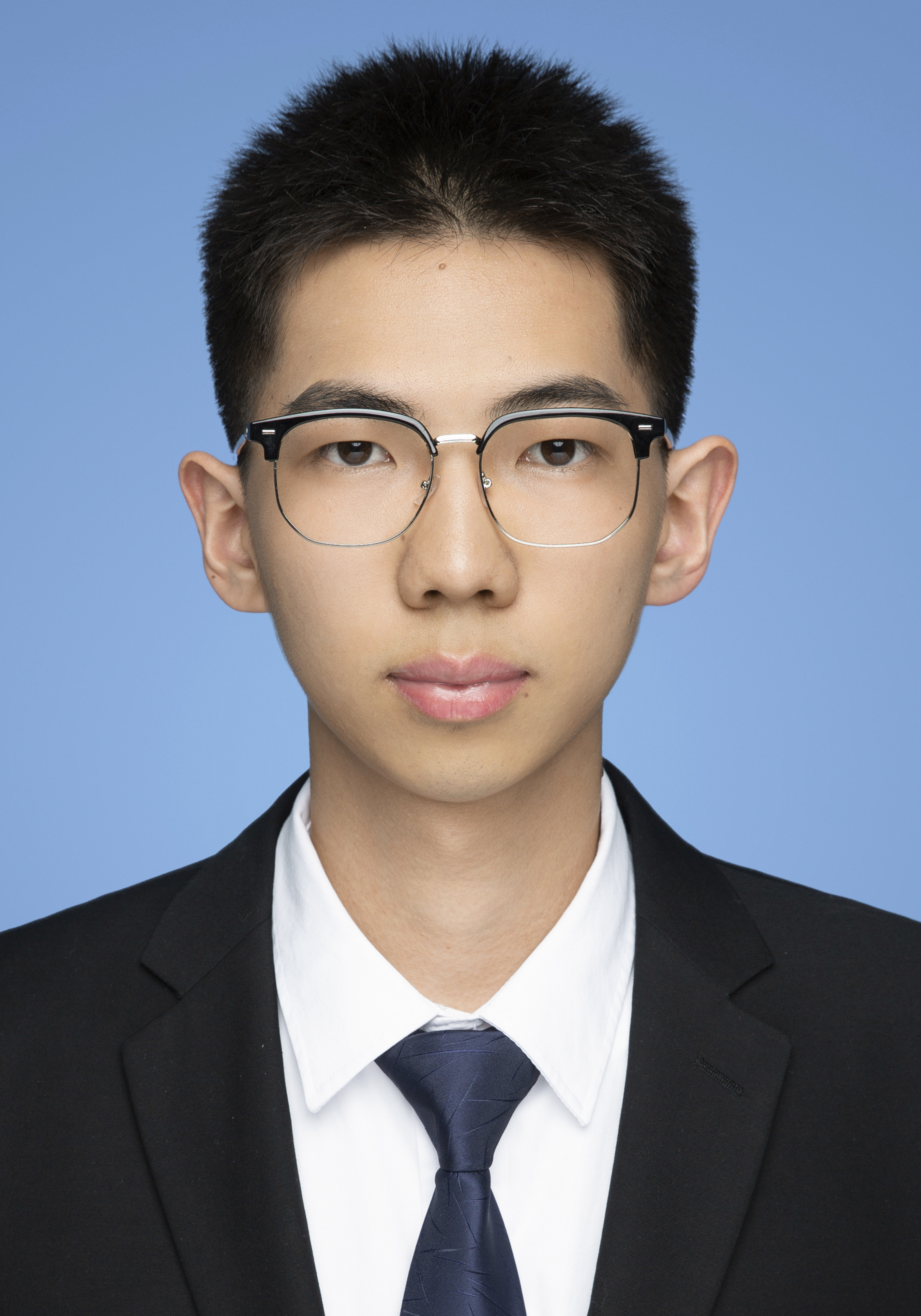}}]{Tan Mo} received the B.S. degree in automation from University of Science and Technology Beijing, Beijing, China, in 2024. He is currently pursuing the M.S. degree in Central South University. 

His main research interests include infrared thermal imaging, temperature compensation, and image processing.

\end{IEEEbiography}

\begin{IEEEbiography}[{\includegraphics[width=1in,height=1.25in,clip,keepaspectratio]{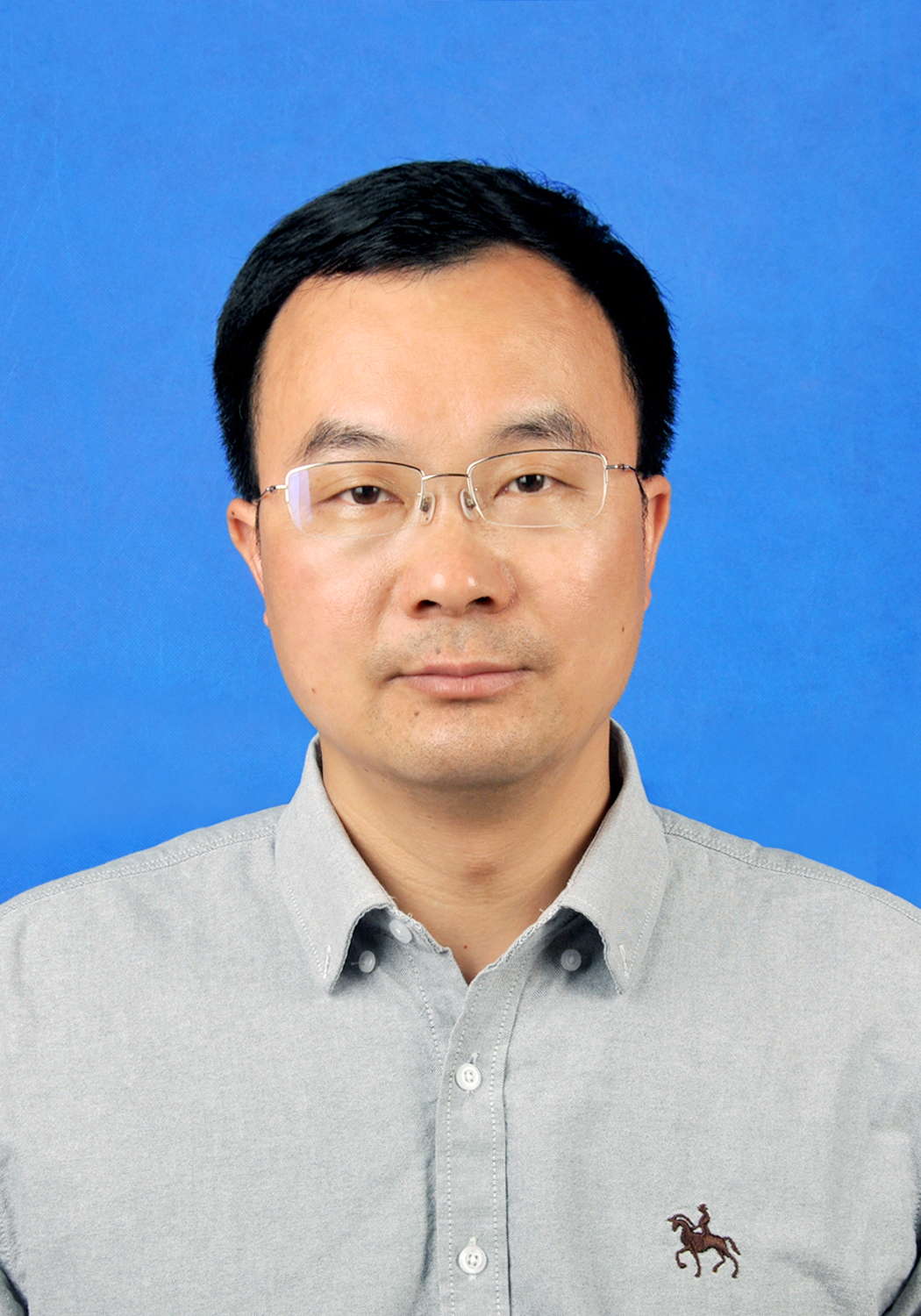}}]{Zhaohui Jiang} (M'19)
received an M. Eng. degree in Automatic Control Engineering and a Ph.D. degree in Control Science and Engineering from Central South University, China in 2006 and 2011, respectively. 

He is currently a professor at Central South University. His research interests include detection technology and automatic equipment, image processing, industrial virtual reality (VR), modeling and optimal control of complex industrial processes.
\end{IEEEbiography}

\begin{IEEEbiography}[{\includegraphics[width=1in,height=1.25in,clip,keepaspectratio]{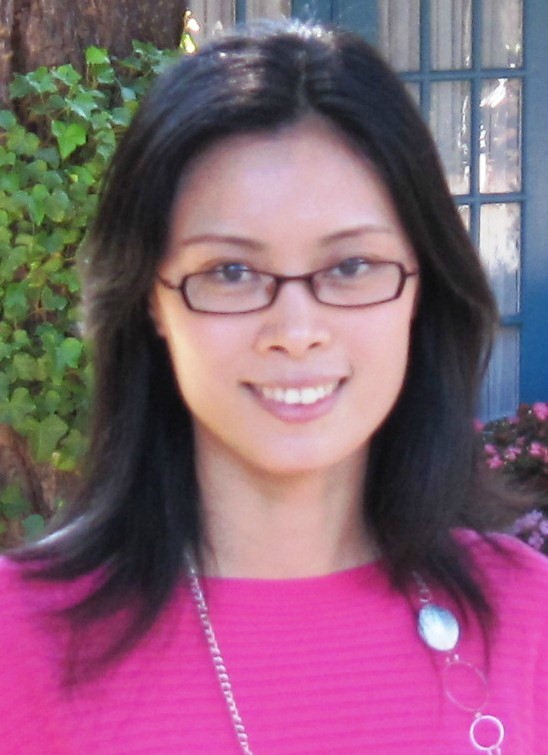}}]{Yuxia Duan} 
received her Ph.D. degree in electrical engineering from Université Laval, Québec City, Canada, in 2014. She is an Associate Professor at Central South University. Her research interests include signal processing for infrared thermography, vision system for industrial inspection, and machine learning target detection using infrared imaging systems. She published over 40 papers in peer-review journals and international conferences.
\end{IEEEbiography}

\begin{IEEEbiography}[{\includegraphics[width=1in,height=1.25in,clip,keepaspectratio]{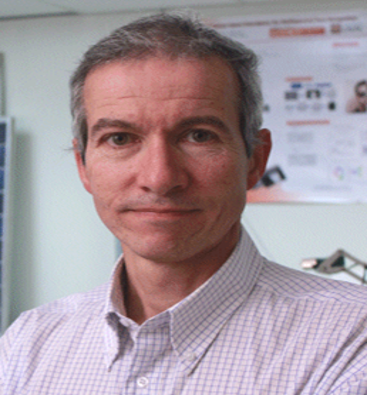}}]{Xavier Maldague} 
has been a full professor at the Department of Electrical and Computing Engineering of Université Laval, Québec City, Canada, since 1989, and the Head of the Department from 2003 to 2008. He has trained over 50 graduate students (M.Sc. and Ph.D.) and has more than 300 publications. His research interests are in infrared thermography, NonDestructive Evaluation (NDE) techniques and vision / digital systems for industrial inspection. He holds the Tier 1 Canada Research Chair in Infrared Vision. He chairs the Quantitative Infrared Thermography (QIRT) Council (since 2004). He is an Honorary Fellow of the Indian Society of Nondestructive Testing, a fellow of the Canadian Engineering Institute, American Society of NonDestructive Testing, Alexandervon Humbolt Foundation (Germany).
\end{IEEEbiography}

\begin{IEEEbiography}[{\includegraphics[width=1in,height=1.25in,clip,keepaspectratio]{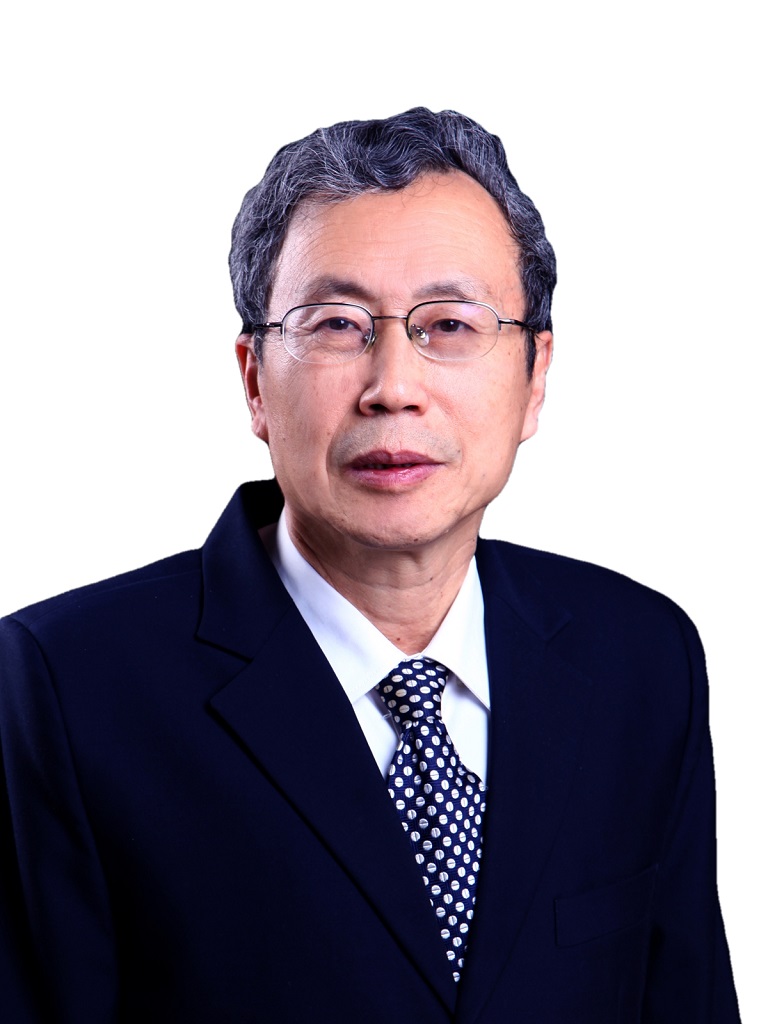}}]{Weihua Gui}
received the degree of the B.Eng. in Electrical Engineering and the M.Eng. in Automatic Control Engineering from Central South University, China in 1976 and 1981, respectively. From 1986 to 1988, he was a visiting scholar at University of Duisburg-Essen, Germany. 

He has been a full professor in Central South University since 1991, and he is an Academician of Chinese Academy of Engineering. His main research interests are in modeling and optimal control of complex industrial process, distributed robust control, and fault diagnoses.
\end{IEEEbiography}
%
%
%





\end{document}